\documentclass[aps,prl,twocolumn,floatfix]{revtex4}
\usepackage{graphicx}
\usepackage{dcolumn}
\usepackage{bm}
\usepackage{amsmath}
\usepackage{epstopdf}
\usepackage{hyperref}

\usepackage{epstopdf, epsfig}
\usepackage{textcomp}
\usepackage{color}

\newcommand{\be}{\begin{equation}}
\newcommand{\ee}{\end{equation}}

\usepackage{blindtext}
\usepackage{lettrine}

\begin{document}

\title{Heat transport scaling and transition in geostrophic rotating convection with varying aspect ratio}
%\title{Heat transport in rapidly rotating Rayleigh-B\'enard convection with varying aspect ratio: scaling and transition}

%\author{Hao-Yuan Lu$^{1}$}
%\author{Guang-Yu Ding$^{2,3}$}
%\author{Guang-Yu Ding$^{2,3 \ref{note1}}$}

\author{Hao-Yuan Lu$^{1}$\footnote{\label{note1}These authors contributed equally to this study}} 
\author{Guang-Yu Ding$^{2,3 \text{\ref{note1}}}$}
\author{Jun-Qiang Shi$^{1}$}
\author{Ke-Qing Xia$^{3,2}$}
\email{xiakq@sustech.edu.cn}
\author{Jin-Qiang Zhong$^{1}$}
\email{jinqiang@tongji.edu.cn}
%\affiliation{$^{1}$School of Physics Science and Engineering, Tongji University, Shanghai 200092, China}
%\affiliation{$^{2}$Department of Physics, The Chinese University of Hong Kong, Shatin, Hong Kong, China}
%\affiliation{$^{3}$Center for Complex Flows and Soft Matter Research and Department of Mechanics and Aerospace Engineering, Southern University of Science and Technology, Shenzhen 518055, China}

\affiliation{$^{1}$School of Physics Science and Engineering, Tongji University, Shanghai 200092, China \\
$^{2}$Department of Physics, The Chinese University of Hong Kong, Shatin, Hong Kong, China\\
$^{3}$Center for Complex Flows and Soft Matter Research and Department of Mechanics and Aerospace Engineering, Southern University of Science and Technology, Shenzhen 518055, China}

\date{\today}

\begin{abstract}
We present high-precision experimental and numerical studies of the Nusselt number $\mathrm{Nu}$ as functions of the Rayleigh number $\mathrm{Ra}$ in geostrophic rotating convection with domain aspect ratio $\Gamma$ varying from 0.4 to 3.8 and the Ekman number $\mathrm{Ek}$ from $2.0{\times}10^{-7}$ to $2.7{\times}10^{-5}$.  
%We determine the power-law scaling of ${Nu}{\sim}{Ra}^{\gamma}$ in rotation-dominated convection, and report that the scaling exponent $\gamma$ increases with increasing $\Gamma$. 
The heat-transport data $\mathrm{Nu(Ra)}$ reveal a gradual transition from buoyancy-dominated to geostrophic convection at large $\mathrm{Ek}$, whereas the transition becomes sharp with decreasing $\mathrm{Ek}$. 
We determine the power-law scaling of $\mathrm{Nu}{\sim}\mathrm{Ra}^{\gamma}$, and show that the boundary flows give rise to pronounced enhancement of $\mathrm{Nu}$ in a broad range of the geostrophic regime, leading to reduction of the scaling exponent $\gamma$ in small $\Gamma$ cells. The present work provides new insight into the heat-transport scaling in geostrophic convection and may explain the discrepancies observed in previous studies.
\end{abstract}

%Rotating convection has been of interest for decades, yet there exists no generally accepted law of heat transport in the flow regime of geostrophic turbulence. Here we present high-precision experimental measurements and numerical simulations of the Nusselt number $Nu$ as functions of the Rayleigh number $Ra$ in rapidly rotating convection in cells with varying aspect ratio $\Gamma$ over a wide range of Ekman number $2.0{\times}10^{-7}{\le}Ek{\le}2.7{\times}10^{-5}$.  Our results reveal a steep scaling, geostrophic regime with ${Nu}{\sim}({Ra/Ra}_c)^{\gamma}$ when $Ra$ falls below a transitional value $Ra_t$. Unexpectedly, we find both the scaling exponent $\gamma$ and the scaling relationship of $Ra_t(Ek)$ depend strongly on $\Gamma$. It is demonstrated that in the geostrophic regime the boundary flows (BFs) have a significant contribution to the global heat transport. For slender cells with small $\Gamma$ the BFs give rise to the pronounce enhancement in $Nu$ and reduce the exponent $\gamma$. The present work may provide new insight into understanding the discrepancy of heat-transport scaling of geostrophic turbulence observed in previous studies.

\maketitle

Buoyancy-induced convection in the presence of rotation occurs widely in the Earth's liquid core \cite{Ol13}, the outer layer of the Sun \cite{MT09}, and the interior of gaseous planets \cite{Bu94}. Heat transport by turbulent flows in rotating convection is an important process for many astro- and geo-physical systems \cite{J011}, and is relevant to numerous industrial applications \cite{KWO07, OL15}. 
%The global heat transport in rotating convection may exhibit various scaling behaviors depending on the flow regimes defined by dimensionless parameters \cite{KSNHA09, JKRV12, KSA12, JRGK12, SLJVCRKA14, EN14, CSRGKA15, Ec15, KOPVL16, CAJK18}, including the Rayleigh number $Ra$ and the Ekman number $Ek$ that characterize buoyancy and rotation, respectively.    
Much of the previous studies has focused on the weak rotation regime \cite{Ro69, LE97, KCG06, ZSCVLA09, SZCAL09, ZA10, NBS10, WWA15} in which $\mathrm{Ra}$ is far above the onset of convection $\mathrm{Ra}_c{=}C\mathrm{Ek}^{-4/3}$ \cite{NB65}, where the Rayleigh number $\mathrm{Ra}$ and the Ekman number $\mathrm{Ek}$ characterize buoyancy and rotation, respectively. In this buoyancy-dominated flow regime the thermal boundary layers (BL) remain the main throttle to the heat transfer. When the rotation rate $\Omega$ increases and the ratio $\mathrm{Ra/Ra}_c$ falls below a transitional value, flow transition occurs from BL controlled convection to geostrophic convection where the local balance of Coriolis force and pressure gradient dominates the bulk flows \cite{JRGK12, EN14}. Although geostrophic convection possesses many important features of astro- and geo-physical flows \cite{KSNHA09, CSRGKA15, ACCJKNSS15}, it has been a challenge to access this flow regime, particularly for high-resolution measurements of the heat transport when both the turbulent thermal forcing and strong rotations ($\mathrm{Ek}{\sim}10^{-7}$) are present \cite{EN14, CSRGKA15, ACCJKNSS15, CAJK18}. 

To achieve a wide parameter range of low $\mathrm{Ek}$, recent experimental and numerical studies \cite{CSRGKA15, KOPVL16, WGMCCK20} have used convection cells with small aspect ratios $\Gamma{=}D/H$ ($D$ and $H$ being the horizontal and vertical scale of the fluid domain), in the hope that measurements in these small-$\Gamma$ domains still provide adequate sampling of the flow structures, since their horizontal scale ($l{\sim}\mathrm{Ek}^{1/3}H$) decreases with decreasing $\mathrm{Ek}$ \cite{Ch61}.  
%Since with decreasing $Ek$ the horizontal scale of the flows decreases as $l{\sim}{Ek}^{1/3}H$ \cite{Ch61},  convection samples with small $\Gamma$ may still ensure adequate sampling of the flow structures. 
However, it remains an unanswered question whether the scaling of heat-transport determined in small $\Gamma$ convection domains can be extrapolated to laterally extended and even unbounded systems.       

The fluid dynamics of rotating, buoyancy-driven flows is often studied by a paradigmatic model, the rotating Rayleigh-B\'enard convection (RBC), i.e., a fluid layer being heated from below and rotated about a vertical axis. 
% i.e., a fluid layer being heated from below and rotated about a vertical axis.  
%In this convection state where the local balance of Coriolis force and the pressure gradient determines the interior flows, the heat transport is significantly suppressed, and 
In the geostrophic regime, the heat transport by rotating RBC, expressed by the Nusselt number $\mathrm{Nu}$, exhibits a steep power-law scaling $\mathrm{Nu}{\sim}(\mathrm{Ra/Ra}_c)^{\gamma}$ \cite{Onset}. Asymptotic theory predicted that in this flow regime $\mathrm{Nu}{=}(\mathrm{Ra/Ra}_c)^{3/2}$, based on the argument that the scaling exponent should be independent of the fluid dissipation properties \cite{JKRV12}. Experimental results suggested $\mathrm{Nu}{\sim}(\mathrm{Ra/Ra}_c)^{3}$, which was interpreted through the BL crossing hypotheses \cite{KSA12}. The $\gamma{=}3$ scaling was found in numerical simulations of the asymptotic theory when the effect of Ekman transport through non-slip boundaries was considered \cite{SLJVCRKA14}. However, recent experiment revealed $1.2{<}\gamma{<}1.6$ \cite{EN14}. Despite the large amount of experimental and numerical studies, there exists hitherto no generally accepted law of heat transport in the geostrophic convection regime. We note that these studies are conducted in domains with different aspect ratios under different boundary conditions.

\begin{figure}
\includegraphics[width=0.48\textwidth]{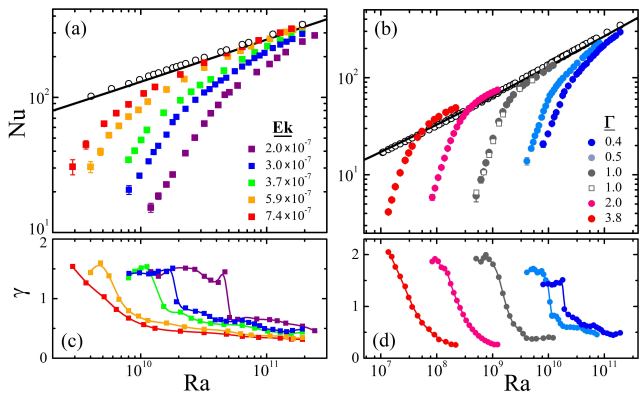}

\caption{(a, b) The Nusselt number as functions of the $Ra$ on logarithmic scales. (a) $\mathrm{Nu(Ra)}$ are measured in the $\Gamma{=}0.4$ cell with various $\Omega$, and (b) with a fixed rotation rate $\Omega{=}3.14$ rad/s but varying $\Gamma$. 
Results in (b) are for $\mathrm{Ek}{=}2.7{\times}10^{-5}$ (red), $7.4{\times}10^{-6}$ (plum), $1.9{\times}10^{-6}$ (gray), $4.6{\times}10^{-7}$ (light blue) and $3.0{\times}10^{-7}$ (blue). 
Open squares: DNS data for $\Gamma{=}1$ and $\mathrm{Ek}{=}1.9{\times}10^{-6}$. Open circles: data for $\Omega{=}0$. Solid lines: the GL theory \cite{SCL13}. (c, d) The local exponent $\gamma{=}d(ln\mathrm{Nu})/d(ln\mathrm{Ra})$ determined from a power-law fit of $\mathrm{Nu(Ra)}$ over various restricted ranges. Symbols are defined in (a) and (b), respectively. Solid curves are guides to the eye.}
\label{fig:1}
\end{figure} 

In this Letter we demonstrate both experimentally and numerically that the scaling properties of heat transport in geostrophic rotating RBC depends sensitively on the aspect ratio $\Gamma$ of the fluid domain. 
Remarkably, we report that the boundary flows in the sidewall region gives rise to pronounced heat-transport enhancement, leading to a much slower scaling of $\mathrm{Nu(Ra)}$ in slender convection cells.   
Our convection apparatus was designed for high-precision heat transport measurement in rotating RBC \cite{SLZ16, ZLW17, SLDZ20}. We used cylindrical cells that had copper top and bottom plates and Plexiglas sidewalls with an inner diameter $D{=}240$ mm and various heights ($H{=}$63, 120, 240, 480, 600 mm), yielding $\Gamma{=}$3.8, 2.0, 1.0, 0.5 and 0.4, respectively. 
Deionized water at a mean temperature of $40.00^{\circ}$C was used as the working fluid. Measurements of $\mathrm{Nu}{=}qH/{\lambda}{\Delta}T$ were taken with rotation rates $\Omega$ up to 4.71 rad/s and various applied temperature differences ${\Delta}T$. The parameter range for $\mathrm{Ra}{=}{\alpha}g{\Delta}TH^3/{\kappa}{\nu}$ and $\mathrm{Ek}{=}\nu/2{\Omega}H^2$ was $1.4{\times}10^7{\le}\mathrm{Ra}{\le}2.9{\times}10^{11}$ and $2.0{\times}10^{-7}{\le}\mathrm{Ek}{\le}2.7{\times}10^{-5}$. Here $q$ is the heat-current density, $g$ is the gravitational acceleration, $\nu, \kappa, \lambda$ are the fluid kinematic viscosity, thermal diffusivity and conductivity, respectively. Thus the reduced Rayleigh number spans $1.3{\le}\mathrm{Ra/Ra}_c{\le}343$. 
% The Froude number $Fr{=}{\Omega}^2D/2g$ spanned $0{\le}Fr{\le}0.27$. 
In the low-$\mathrm{Ek}$ regime ($\mathrm{Ek}{<}10^{-6}$), the present study extended the measurement range, reducing $\mathrm{Ra/Ra_c}$ by about half a decade compared to earlier measurements in water \cite{CSRGKA15, WGMCCK20}. We refer to the rotation-dominated flow regime as \emph{geostrophic convection} where $\mathrm{Ra}$ is above the convective onset but below a critical value $\mathrm{Ra}_t$ (defined below) for heat-transport scaling transition, as in this region the geostrophic balance holds \cite{GT}.
%Previous studies of flow morphology in rapidly rotating RBC revealed multiple behavioral regimes (e.g. geostrophic turbulence, plumes, columnar and cellular) \cite{JRGK12}. Here, based on the scaling properties of $Nu(Ra)$ we refer to the rotation-dominated regime as \emph{geostrophic convection} where $Ra$ is below a transitional value $Ra_t$ for regime transition, in viewing that geostrophic balance holds well in this flow regime \cite{EN14} . 
%With increasing $\Omega$, the convection flow in rotating RBC changes morphology, resulting in multiple behavioral regimes (e.g. geostrophic turbulence, plumes, columnar) according to Ref. \cite{JRGK12}. In Ref. \cite{EN14}, the regime of geostrophic turbulence is determined through the global heat-transport properties, which covers the multiple flow domains within $3Ra_c{\le}Ra{\le}Ra_t$. We adopt the definition of \cite{EN14} in discussions of the geostrophic regime.
We also made direct numerical simulations (DNS) that solved the Navier-Stokes equations in cylindrical cells with non-slip boundaries, using the multiple-resolution version of the {\it{CUPS}} code \cite{KX13, CDX18, Paper1}. The numerical and experimental data covered different and overlapping parameter ranges and complemented each other; and where their parameter ranges overlapped, the corresponding data were in close agreements (see Supplemental Material \cite{SM} for detailed experimental and numerical methods).
%\cite{ZSL15, SLZ16, ZLW17, DLYZ19, SLDZ20}.
%\cite{KX13, KSX14, CDX18, Paper1}.
% The simulation was performed in cylindrical cells with ${Ek}{=}1.9{\times}10^{-6}$ and with various aspect ratios ($\Gamma{=}0.5, 1.0, 2.0$)

\begin{figure}
\includegraphics[width=0.48\textwidth]{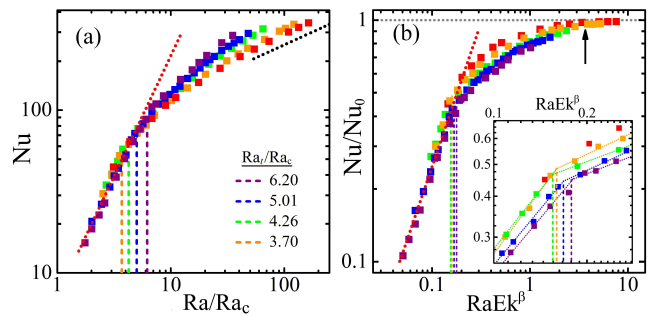}
\caption{(a) $\mathrm{Nu}$ as a function of $\mathrm{Ra/Ra}_c$ for various $\mathrm{Ek}$ with $\Gamma{=}0.4$. Symbols are defined in Fig.\ 1a. The vertical dashed lines denote the first transition values $\mathrm{Ra}_{t}/\mathrm{Ra}_c$ for $\mathrm{Ek}{=}2.0{\times}10^{-7}$ (purple), $3.0{\times}10^{-7}$(blue), $3.7{\times}10^{-7}$ (green) and $5.9{\times}10^{-7}$ (orange). The red dotted line represents the power-law fit to the data, $\mathrm{Nu}{\sim}(\mathrm{Ra/Ra}_c)^{1.48}$ for the range $\mathrm{Ra}{\le}\mathrm{Ra}_{t}(Ek)$. The black dotted line indicates the non-rotating scaling $\mathrm{Nu_0}{\sim}\mathrm{Ra}^{0.317}$. (b) The ratio $\mathrm{Nu/Nu}_0$ as a function of $\mathrm{RaEk}^{\beta}$ with $\mathrm{Nu}_0{=}0.0921\mathrm{Ra}^{0.317}$ and the exponent $\beta{=}1.70$ for $\Gamma{=}0.4$. The red dotted line represents the power-law fit to the data, $\mathrm{Nu/Nu}_0{\sim}(\mathrm{RaEk}^{\beta})^{1.19}$ for $\mathrm{Ra}{\le}\mathrm{Ra}_{t}$. The vertical dashed lines denote the first transition $\mathrm{Ra}_t\mathrm{Ek}^{\beta}{=}0.17{\pm}0.01$. The arrow indicates approximately the second transition $\mathrm{Ra}_{t2}\mathrm{Ek}^{\beta}{=}3.4{\pm}0.4$. Inset: an expanded view in the vicinity of the first transition $\mathrm{Ra}_t\mathrm{Ek}^{\beta}$. For each $\mathrm{Ek}$, $\mathrm{Ra}_t$ is determined by the interaction of the two locally fitted power-law lines.}
\label{fig:2}
\end{figure} 

Our measurements of $\mathrm{Nu}$ with $\Omega{=}0$ suggest a heat-transport scaling $\mathrm{Nu}_0{\sim}\mathrm{Ra}^{\gamma_0}$ that agrees, within estimated systematic errors of about $2\%$, with previous studies, and with the Grossman-Lohse (GL) theory \cite{SCL13} as shown in Figs.\ 1a and 1b \cite{SM}. Data of $\mathrm{Nu}$ obtained from the $\Gamma{=}0.4$ cell with various rotation rates $\Omega$ are presented as functions of $\mathrm{Ra}$ in Fig.\ 1a. For each $\mathrm{Ek}$ we see that with increasing $\mathrm{Ra}$ the Nusselt number first rises steeply in the geostrophic convection regime. For strong enough buoyancy forcing the measured $\mathrm{Nu(Ra)}$ conform to the non-rotating behavior.  Similar transitional behavior is observed in Fig.\ 1b which shows results of $\mathrm{Nu(Ra)}$ from five cells with different $\Gamma$ at the same rotation rate $\Omega{=}3.14$ rad/s. A striking feature revealed in Figs.\ 1a and 1b is an Ek-dependent transitional behavior of $\mathrm{Nu(Ra)}$ from geostrophic convection to buoyancy-dominated convection, i.e., the transition is gradual at high $\mathrm{Ek}$ but becomes increasingly sharper at low $\mathrm{Ek}$.  
%We find the threshold value $Ek_c{\approx}5.9{\times}10^{-7}$ for $\Gamma{=}0.4$ (Fig.\ 1a), which appears to depend on $\Gamma$ (Fig.\ 1b). With low $Ek$ (${Ek}{\le}Ek_c$) the heat transport data exhibit in the geostrophic convection regime a steep power-law scaling that terminates abruptly at a certain transitional value of $Ra$.  
The different transitional properties of $\mathrm{Nu(Ra)}$ can be seen more clearly in Figs.\ 1c and 1d, where we show the local exponent, $\gamma{=}d(ln\mathrm{Nu})/d(ln\mathrm{Ra})$, as functions of $\mathrm{Ra}$ for various $\mathrm{Ek}$ and $\Gamma$. For $\Gamma{=}0.4$ (Fig.\ 1c), one can divide roughly the Ek-range into two domains: a gradual-transition domain with $\mathrm{Ek}{\ge}5.9{\times}10^{-7}$ and a sharp-transition domain with $\mathrm{Ek}{<}5.9{\times}10^{-7}$. Similar transitional behavior is observed with varying aspect ratios in Fig.\ 1d.    
%Results of $\gamma$ are plotted in Figs.\ 1c and 1d, which provide a quantitative representation of the transitional properties of ${Nu(Ra)}$. 
It remains a phenomenon of interest but unexplained to us that the transition between the BL-controlled and the geostrophic convection is sudden for low $\mathrm{Ek}$, but becomes less abrupt and eventually smooth at high $\mathrm{Ek}$. Since in Fig.\ 1 data sets with low $\mathrm{Ek}$ have higher values of $\mathrm{Ra}$, for which various coherent turbulent structures arise in the flow field under rotations \cite{JRGK12, SLJVCRKA14, CSRGKA15}, we speculate that the corresponding sharper transitions may be related to the different properties of turbulent structures that modify the heat-transport scaling, analogous to previous findings in weakly rotating RBC \cite{SZCAL09, WWA15}. 
%The gradual transition found in a high-$Ek$ (small $Ra$) range, however, may be owing to the lack of turbulence structures and the weakening of the rotational constraints. 
Further studies, both experimental and numerical, are needed to substantiate this argument. 

% Since in Fig.\ 1 data curves of low $Ek$ correspond to a parameter range of high-${Ra}$ convection, where various coherent turbulent structures arise under the constraints of strong rotations \cite{JRGK12, SLJVCRKA14, CSRGKA15}, we speculate that it is the sudden changes in the properties (e.g. the length scale, flow symmetry) of these turbulent structures that lead to the sharp transition, as suggested in previous studies of weakly rotating RBC \cite{SZCAL09, WWA15}. In a high-$Ek$ range where $Ra$ is relatively small, however, the observed gradual transition in ${Nu(Ra)}$ may be owing to the lack of turbulence structures and the weakening of the rotational constraints.      

\begin{table}
\begin{ruledtabular}
\begin{tabular}{ccccccccc}
 $\Gamma$&$10^6\mathrm{Ek}$&$\gamma$&$\gamma_0$&$\beta$&$\mathrm{Ra}_{t}\mathrm{Ek}^{\beta}$\\ \hline
0.4\scriptsize{(EXP)}&0.30&$1.48{\pm}0.06$&0.317&1.70&0.17\\
0.5\scriptsize{(EXP)}&0.46&$1.65{\pm}0.03$&0.316&1.65&0.36\\
0.5\scriptsize{(DNS)}&1.9&$1.42{\pm}0.06$&&&\\
1.0\scriptsize{(EXP)}&1.9&$1.82{\pm}0.06$&0.303&1.60&1.01\\
1.0\scriptsize{(DNS)}&1.9&$1.77{\pm}0.05$&&&\\
2.0\scriptsize{(EXP)}&7.4&$1.81{\pm}0.05$&0.302&1.60&1.18\\
2.0\scriptsize{(DNS)}&1.9&$2.04{\pm}0.05$&&&\\
\end{tabular}
\end{ruledtabular}
\caption{\label{tab:table1}
Experimental (EXP) and numerical (DNS) results for the scaling exponents $\gamma, {\gamma_0}$, $\beta$ and the transition value $\mathrm{Ra}_{t}\mathrm{Ek}^{\beta}$.}
\end{table}

In Fig.\ 2 we examine the scaling properties of $\mathrm{Nu}$ measured in the $\Gamma{=}0.4$ cell. Figure 2a shows $\mathrm{Nu}$ as a function of $\mathrm{Ra/Ra}_c$ \cite{Rac} for various $\mathrm{Ek}$. The data collapse approximately in the geostrophic convection regime where $\mathrm{Ra}$ is below an Ek-dependent transition value $\mathrm{Ra}_t$. $\mathrm{Ra}_t$ is determined as the upper bound of the steep power-law scaling of $\mathrm{Nu(Ra)}$ (shown in the inset of Fig.\ 2b). 
%Fig.\ 2a shows that ${Ra_t}/Ra_c$ increases apparently with decrease in $Ek$. 
Linear regression of the data with $\mathrm{Ra}{\le}\mathrm{Ra}_t$ in the log-log plot suggests a power law $\mathrm{Nu}{\sim}(\mathrm{Ra/Ra}_c)^{\gamma}$, with the fitted exponent $\gamma$ and its statistical error given in Table 1. The range of the power-law dependence expands as $\mathrm{Ek}$ decreases, since $\mathrm{Ra}_t/\mathrm{Ra}_c$ increases apparently with decreasing $\mathrm{Ek}$. 
We note that for a higher $\mathrm{Ek}$ fewer data points are available in the geostrophic convection regime. The power-law fitting here is thus in fact restricted in a relatively small range of $\mathrm{Ek}$. 
For $\mathrm{Ra}{>}\mathrm{Ra}_t$, $\mathrm{Nu}(\mathrm{Ra/Ra}_c)$ becomes dependent on $\mathrm{Ek}$ with a greater value for a lower $\mathrm{Ek}$. The spread of the data in this flow regime was reported and ascribed to the Ekman pumping effect \cite{SLJVCRKA14, PJMS16, JACKMSV16} that enhances $\mathrm{Nu}$ with its strength depending on $\mathrm{Ek}$. One expects that in the limit of large $\mathrm{Ra}$ the heat-transport data approach the non-rotating scaling $\mathrm{Nu}{\sim}\mathrm{Ra}^{\gamma_0}$, as shown in Fig.\ 2a.        

\begin{figure}
\includegraphics[width=0.48\textwidth]{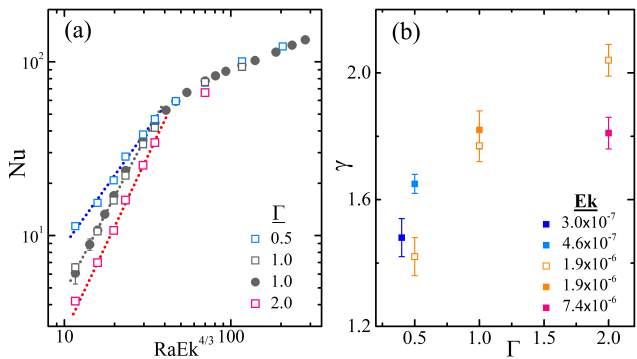}
\caption{(a) $\mathrm{Nu}$ as a function of $\mathrm{RaEk}^{4/3}$. Results for $\mathrm{Ek}{=}1.9{\times}10^{-6}$ with $\Gamma{=}0.5$ (light blue), 1.0 (gray) and 2.0 (plum). The dashed lines are power-law fits to the data in the steep-scaling regime. (b) The exponent $\gamma$ as a function of $\Gamma$ for various $\mathrm{Ek}$. Filled symbols: Experimental data. Open symbols: DNS data.}
\label{fig:3}
\end{figure} 

Our heat transport data in the non-rotating regime $\mathrm{Nu}_0{\sim}\mathrm{Ra}^{\gamma_0}$, and in the geostrophic convection regime $\mathrm{Nu}{\sim}(\mathrm{Ra/Ra}_c)^{\gamma}$ suggest that one may rescale the data as $\mathrm{Nu/Nu}_0{\sim}(\mathrm{RaEk}^{\beta})^{\gamma{-}\gamma_0}$ for geostrophic convection, with the exponent $\beta{\equiv}4{\gamma}/3({\gamma}{-}{\gamma_0})$ \cite{EN14, CSRGKA15}. Figure 2b plots $\mathrm{Nu/Nu}_0$ as a function of $\mathrm{RaEk}^{\beta}$ for $\Gamma{=}0.4$. We see that indeed data with $\mathrm{Ra}{\le}\mathrm{Ra}_t$ collapse onto the predicted power-law for various $\mathrm{Ek}$. Interestingly, regardless of $\mathrm{Ek}$ the transitional values $\mathrm{Ra}_t\mathrm{Ek}^{\beta}$ converge approximately into the same location, and suggest a relationship of regime transition for $\Gamma{=}0.4$: $\mathrm{Ra}_t{\sim}0.17\mathrm{Ek}^{-\beta}$. Figure 2b also reveals a second transition at $\mathrm{Ra}_{t2}{\sim}3.4\mathrm{Ek}^{-\beta}$: $\mathrm{Nu}$ approaches the non-rotating value $\mathrm{Nu}_0$ for $\mathrm{Ra}{\ge}\mathrm{Ra}_{t2}$. The values of the exponents $\beta$ and the transitional values of $\mathrm{Ra}_t\mathrm{Ek}^{\beta}$ for various $\Gamma$ and $\mathrm{Ek}$ are listed in Table 1. In the crossover regime $0.17{\le}\mathrm{RaEk}^{\beta}{\le}3.4$ we see that for a given $\mathrm{RaEk}^{\beta}$, $\mathrm{Nu/Nu}_0$ decreases with decreasing $\mathrm{Ek}$. These data imply that the asymptotic behavior of rotating convection predicted in \cite{JACKMSV16}, i.e., the scaled heat-transport data become independent of $\mathrm{Ek}$, is yet to be observed at even lower $\mathrm{Ek}$. 

%No clear evidence is observed for data convergence towards the asymptotic behavior \cite{JACKMSV16} in the range of $Ek$ studied. 
% In this weakly rotating regime, we do not observe the heat-transport enhancement with ${Nu}{>}{Nu}_0$ in Fig.\ 2b (or Fig.\ 1b), since the chosen ${Ra}$ here are far beyond the parameter range where the Ekman-pumping process acts dominantly to increase $Nu$ [ref]. 
% converge into the same value of  $Ra_t{Ek}^{\beta}{=}0.17$ where the steep scaling regime is truncated for various $Ek$. 

\begin{figure}
\includegraphics[width=0.5\textwidth]{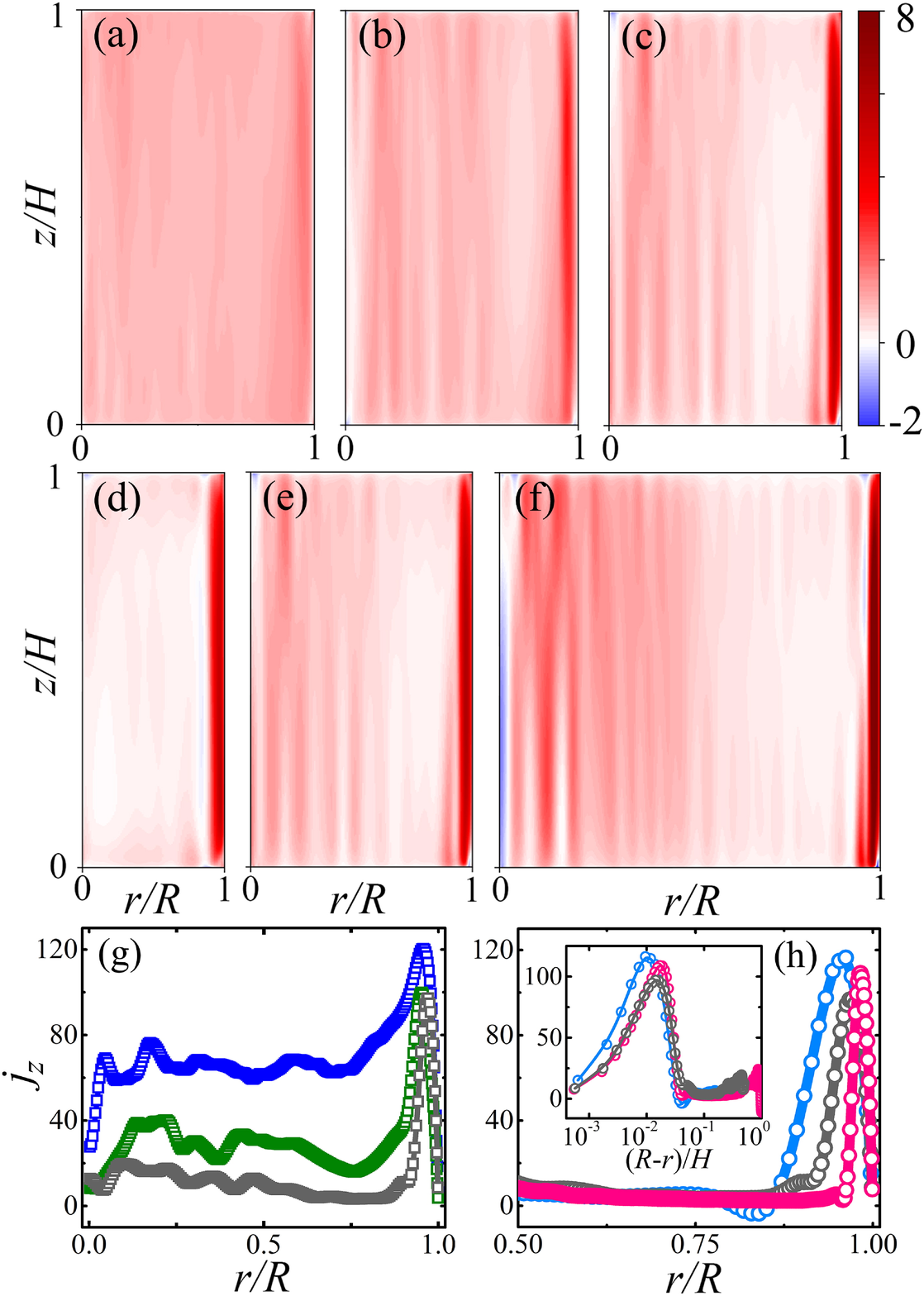}
\caption{(a-f) DNS data for the normalized time- and azimuthally-averaged vertical convective heat flux $J_z(r, z){=}\langle{\mathcal{J}_z(t, r, \phi, z)}\rangle_{t, \phi}/\mathrm{Nu}$ \cite{SM}. (a-c) Results for $\Gamma{=}1$, $\mathrm{Ek}{=}1.9{\times}10^{-6}$ and for $10^{-9}\mathrm{Ra}{=}$3.07, 1.30 and 0.87, respectively. The corresponding radial profiles of the absolute (unnormalized) heat flux, $j_z(r){=}\langle{\mathcal{J}_z(t, r, \phi, z{=}H/2)}\rangle_{t, \phi}$, are shown as the blue, green and gray curves in (g). (d-f) $J_z(r, z)$ for $\mathrm{Ra}{=}8.7{\times}10^8$, $\mathrm{Ek}{=}1.9{\times}10^{-6}$ with $\Gamma{=}$ 0.5, 1.0 and 2.0, respectively. Their corresponding radial profiles $j_z(r)$ are shown as the light blue, gray and plum curves in (h). Inset: $j_z(r)$ as functions of $(R{-}r)/H$ on a logarithmic scale.}
\label{fig:4}
\end{figure}

The heat-transport data from the $\Gamma{=}0.4$ cell, which cover a relatively small range of $\mathrm{Ek}$, are compatible with the power law scaling of $\mathrm{Nu}{\sim}\mathrm{Ra}^{1.48}$ in the geostrophic convection regime (Fig.\ 2). However, results of $\mathrm{Nu(Ra)}$ over a wider parameter range, depicted in Fig.\ 1b and 1d, suggest that the exponent $\gamma$ is dependent on both of the control parameters $\Gamma$ and $\mathrm{Ek}$. Surprisingly, we find that $\gamma$ decreases in slender cells with smaller $\Gamma$ that span a parameter range of lower $\mathrm{Ek}$ (Table 1), which appears to be contrary to what has been previously observed that the heat-transport scaling becomes steeper with decreasing $\mathrm{Ek}$ \cite{CSRGKA15, JACKMSV16, CAJK18}.
To understand the $\mathrm{Ek}$- and $\Gamma$-dependence of $\gamma$ in the geostrophic convection regime, we present in Fig.\ 3a both experimental and numerical results of $\mathrm{Nu}(\mathrm{RaEk}^{4/3})$ for three values of $\Gamma$ but with fixed $\mathrm{Ek}{=}1.9{\times}10^{-6}$. One sees that in the geostrophic convection regime with $\mathrm{Ra}{\le}\mathrm{Ra}_t$, $\mathrm{Nu(Ra)}$ exhibits a steeper power-law for a larger $\Gamma$. With increasing $\mathrm{Ra/Ra}_c$ the three sets of data converge approximately at the same transition point towards the geostrophic turbulence regime, suggesting that the transitional value $Ra_t$ is independent of $\Gamma$. Figure 3b plots the exponent $\gamma$ as a function of $\Gamma$ for various $\mathrm{Ek}$. We find that for a given $\mathrm{Ek}$ (e.g., $1.9{\times}10^{-6}$), $\gamma$ increases strongly with the aspect ratio; whereas for a fixed $\Gamma$, $\gamma$ increases with decreasing $\mathrm{Ek}$. These results also suggest, for the parameter range studied, that the scaling exponent $\gamma$ depends more sensitively on $\Gamma$ than on $\mathrm{Ek}$ for geostrophic convection. 

%one sees the ``crossing" of the data curves in the steep power-law scaling regime, indicating that in the present parameter range the power-law dependence of $Nu$ on $Ra{Ek}^{4/3}$ has a slightly higher power exponent with a smaller pre-factor for lower $Ek$ (thus higher $Ra$).   
%in the power-law dependence of $Nu{\sim}$ on $Ra{Ek}^{4/3}$, for a lower $Ek$ (higher $Ra$). 
%reveal that for a given $\Gamma$ the power-law dependence of ${Nu(Ra{Ek}^{4/3})}$

Figure 3a shows that the different heat-transport scaling with varying $\Gamma$ results in a higher Nusselt number for the slender cell ($\Gamma{=}0.5$), which exceeds the corresponding value for a wide cell ($\Gamma{=}2.0$) by over $150\%$ for the same $\mathrm{Ra}$ near the convection onset. 
The enhancement of $\mathrm{Nu}$ for slender cells is observed in a wide range of geostrophic convection even for $\mathrm{Ra}$ being far above the onset. To understand this phenomenon, we visualize the flow field (Figs.\ 4a-4f ) and show in vertical cross-sections the normalized time- and azimuthally-averaged vertical convective heat flux $J_z(r, z)$ \cite{SM}. Figures 4a-4c present the results for $\Gamma{=}1.0$ with varying $\mathrm{Ra}$ and fixed $\mathrm{Ek}$. For low $\mathrm{Ra}$ one sees clearly that near the sidewall ($r{=}R{\equiv}D/2$) there is a region of large local heat flux, indicating the existence of a boundary flow (BF) structure \cite{WGMCCK20, ZvHWZAEWBS20, EK20}. Such a flow feature is represented by the dominant peak in the radial profiles $j_z(r)$ of the absolute heat flux evaluated in the mid-plane of the cell (Fig.\ 4g) \cite{SM}. With decreasing $\mathrm{Ra}$, while $j_z$ gradually decreases in the bulk region, it remains in a significant value in the sidewall region. Thus the contribution by BF to the global heat-transport enhancement increases with decreasing $\mathrm{Ra}$. 
%CTC structures in (b) and (c)É  

Figures 4d-4f compare the distributions of $J_z(r, z)$ for $\mathrm{RaEk}^{4/3}{=}19.8$ in cells with $\Gamma{=}$0.5, 1.0 and 2.0. We find that although the lateral extend of the fluid layer increases with increasing $\Gamma$, the spatial structures of the boundary flow are similar. This is shown in the expanded view of the radial profiles $j_z(r)$ in the inset of Fig.\ 4h, as the peaks of $j_z(r)$ exhibit a similar structure with approximately the same magnitude and width. It is for this aspect-ratio invariant properties that the BF gives rise to a larger $\mathrm{Nu}$ in slender cells: with a smaller $\Gamma$ the BF occupies a relatively larger volume of the cell (see the radially-scaled plot of $j_z(r)$ in Fig.\ 4h), and makes a greater contribution to the overall enhancement in $\mathrm{Nu}$. 
%Interestingly, we find that the enhanced heat flux in the BF region, albeit with the non-zero local heat flux in the bulk, finally leads to a power-law scaling of $Nu(Ra)$ as seen in Fig.\ 3a.% 
Near the rotation axis ($r{=}0$) $J_z(r, z)$ remains small, indicating that in this regime the centrifugal effect is insignificant for heat transport by the bulk flows, in line with previous studies \cite{HA18, HA19}.    

We have shown that lateral constraint of the flow domain impacts strongly the scaling properties of heat-transport in the geostrophic rotating RBC. It is demonstrated that the local heat flux carried by the boundary flow (BF) makes up a significant portion of the global heat transport in a broad range of geostrophic convection, leading to the unexpected aspect-ratio-dependence of both the scaling exponents and the critical values for regime transitions. We conclude that the scaling relationship of $\mathrm{Nu(Ra)}$ measured in convection cells with finite $\Gamma$ cannot be extrapolated to most large-scale, laterally unbounded geophysical and astrophysical flows, while theories of geostrophic convection that neglect the lateral boundary confinements provide an incomplete description of laboratory experiments. The present study brings new insight into understanding the diverse results of heat-transport scaling obtained from previous experiments and simulations \cite{KSNHA09, JKRV12, KSA12, SLJVCRKA14, EN14,CSRGKA15, KOPVL16}. 
%Finally, our findings that for slender cells the boundary flows produce a larger heat-transfer in rotating convection, as compared to for larger aspect ratio cells, may have practical applications of thermal management, such as the design of heat-exchange cavities for rotary devices and other industry processes.  

This work is supported by the National Science Foundation of China under Grant No. 11772235, a NSFC/RGC Joint Research Grant No. 1561161004 (JQZ) and N\_CUHK437$/$15 (KQX) and by the Hong Kong Research Grants Council under Grant No. 14302317. Computing resources are provided by the SUSTech Center for Computational Science and Engineering. 

%H.-Y. L.  and G.-Y. D. contributed equally to this study.

%\bibliography{refs_all_long}

\begin{thebibliography}{45}
\bibitem[{Ons()}]{Onset}
\bibinfo{note}{Over the range very close to the onset of convection, weakly
  nonlinear theory \cite{CH93} suggests an alternative representation of the
  data: $\mathrm{Nu}{-}1{=}c_1{\epsilon}{+}c_2{\epsilon}^2{+}...$, where
  $\epsilon{=}\mathrm{Ra/Ra}_c{-}1$, $c_1$ and $c_2$ are constants. Here we
  focus on discussions of the power-law scaling of $\mathrm{Nu(Ra/Ra}_c)$ with
  $\mathrm{Ra/Ra}_c$ covering a sufficiently large domain in the geostrophic
  regime.}
\bibitem[{Rac()}]{Rac}
\bibinfo{note}{Per Ref. \cite{NB65}, the coefficient $C$ in
  $\mathrm{Ra}_c{=}C\mathrm{Ek}^{-4/3}$ depends on $\mathrm{Ek}$ according to
  $C{=}8.70{-}9.63\mathrm{Ek}^{1/6}$, with values between 7.78 and 7.95 over
  the range of $\mathrm{Ek}$ in Fig.\ 2. We use the mean $C{=}7.87$ to
  determine $\mathrm{Ra}_c$ for Fig.\ 2}.
\end{thebibliography}

\end{document}